\documentclass[aps,prl,twocolumn, superscriptaddress,showpacs]{revtex4-1}

\usepackage[T1]{fontenc}
\usepackage[latin1]{inputenc}
\usepackage[dvips]{graphicx,color}
\usepackage{rotating}
\usepackage{bm}        
\usepackage{amssymb}   
\usepackage{amsmath} 
\usepackage{braket}
\def\jcp#1#2#3{J.~Chem.~Phys.~{\bf #1},\ #2\ (#3)}

\def\pra#1#2#3{Phys.~Rev.~A~{\bf #1},\ #2\ (#3)}
\def\prl#1#2#3{Phys.~Rev.~Lett.~{\bf #1},\ #2\ (#3)}

\def\rmp#1#2#3{Rev.~Mod.~Phys.~{\bf #1},\ #2\ (#3)}

\def\k1{k_1}
\def\k2{k_2}
\def\q1{q_1}
\def\q2{q_2}

\def\({\left (}
\def\){\right )}
\def\[{\left [}
\def\]{\right ]}

\newcommand{\beq}{\begin{equation}}
\newcommand{\eeq}{\end{equation}}

\begin{document}
\date{\today}
\title{A fundamental limit to the efficiency of spin-exchange optical pumping of $^3$He nuclei}

\author{T. V. Tscherbul}
\affiliation{Harvard-MIT Center for Ultracold Atoms, Cambridge, Massachusetts 02138}
\affiliation{ITAMP,
Harvard-Smithsonian Center for Astrophysics, Cambridge, Massachusetts 02138}\email[]{tshcherb@cfa.harvard.edu}
\author{P. Zhang}
\affiliation{ITAMP,
Harvard-Smithsonian Center for Astrophysics, Cambridge, Massachusetts 02138}
\author{H. R. Sadeghpour}
\affiliation{ITAMP,
Harvard-Smithsonian Center for Astrophysics, Cambridge, Massachusetts 02138}
\author{A. Dalgarno}
\affiliation{Harvard-MIT Center for Ultracold Atoms, Cambridge, Massachusetts 02138}
\affiliation{ITAMP, 
Harvard-Smithsonian Center for Astrophysics, Cambridge, Massachusetts 02138}

\begin{abstract}
We establish the existence of a fundamental limit to the efficiency of spin-exchange optical pumping of $^3$He nuclei by collisions with spin-polarized alkali-metal atoms. Using accurate {\it ab initio} calculations of molecular interactions and scattering properties, we show that the maximum $^3$He spin polarization that can be achieved in spin-exchange collisions with potassium ($^{39}$K) and silver ($^{107}$Ag) atoms is limited by the anisotropic hyperfine interaction. We find that spin exchange in Ag-He collisions occurs much faster than in K-He collisions, suggesting the possibility of using Ag in spin-exchange optical pumping experiments to increase the production rate of hyperpolarized $^3$He. Our analysis indicates that measurements of trap loss rates of $^2S$ atoms in the presence of cold $^3$He gas may be used to probe anisotropic spin-exchange interactions in atom-He collisions. 
\end{abstract}

\maketitle

Spin exchange optical pumping (SEOP) is a well-established experimental technique for the production of hyperpolarized noble gas nuclei with many applications in diverse areas of science and technology, including nuclear physics \cite{NeutronSpinStructure,NuclearPhysics}, magnetic resonance imaging  \cite{RMP,MRI}, magnetometry \cite{Romalis}, and precision measurement \cite{Petukhov,Kimball}. Recently, SEOP has emerged as a new technique for probing nucleon-nucleon interactions \cite{Petukhov,Romalis,Kimball}. In particular, measurements of nuclear spin relaxation in a hyperpolarized gas of $^3$He can be used to constrain spin-dependent interactions between nucleons \cite{Petukhov}. The study  of spin-exchange collisions between $^3$He and alkali-metal atoms can yield insights into anomalous nuclear forces and short-range torsion gravity fields \cite{Kimball}. The spin-exchange collisions are an important source of noise and decoherence in atomic magnetometers based on alkali-metal vapor cells \cite{Romalis}.

The SEOP technique is based on collision-induced transfer of spin polarization from optically pumped alkali-metal atoms to noble-gas atoms (typically $^3$He) \cite{RMP}. While the efficiency of this strategy has been confirmed by numerous experiments \cite{RMP}, recent experimental work has established that the maximum $^3$He spin polarization that can be achieved in SEOP experiments with alkali-metal atoms K and Rb is limited to 80\% by unknown relaxation mechanisms \cite{Walker02,Walker06,Walker10}. Possible explanations include wall collisions, magnetic field gradients, and anisotropic hyperfine interactions in atom-He collisions \cite{Walker06,Walker10}. While the first two mechanisms can be eliminated by an appropriate design of the buffer-gas cell, the third mechanism cannot be avoided. The anisotropic hyperfine interaction couples the nuclear spin of $^3$He with end-over-end rotation of the collision complex, which leads to an irreversible decay of nuclear spin polarization, fundamentally limiting the efficiency of SEOP for any given alkali-metal-He pair. 

Previous theoretical and experimental work provided evidence for a minor, albeit non-negligible role, of anisotropic hyperfine interactions in spin-exchange  collisions \cite{Walter}. Walter {\it et al.} estimated the maximum $^3$He polarization attainable via Rb-He collisions to be $P_\text{max}=0.95$. The accuracy of this estimate was, however, limited by a lack of information on interaction potentials and hyperfine interactions. Walker {\it et al.} have recently measured $P_\text{max} = 0.90\pm 0.11$ for K-$^3$He collisions at 463.15~K \cite{Walker10}. These findings suggest that while the SEOP efficiency might indeed be limited by the anisotropic hyperfine interaction, other mechanisms (such as wall relaxation) cannot be ruled out based on the experimental data alone. A quantitatively accurate theoretical analysis of spin exchange and spin relaxation mechanisms would not only settle this long-standing question, it will also establish a pathway toward systematic improvement of the SEOP technology.

In this Letter, we use accurate {\it ab initio} calculations of molecular interactions in combination with exact quantum scattering methods \cite{pra08,pra09} to quantify the role of the anisotropic hyperfine interaction in spin-exchange collisions of alkali-metal atoms with $^3$He. Using the K-He collision system as a representative example \cite{Walter,Walker06, Walker07,Walker07a,Baranga,Walker10}, we show that the maximum $^3$He spin polarization attainable in SEOP experiments with K atoms is limited by the anisotropic hyperfine interaction.  
Our results are in quantitative agreement with recent experimental measurements of frequency shift enhancement factors \cite{Walker05} and rate constants for spin exchange in K-$^3$He collisions \cite{Walker07,Walker10}. In addition, we show that spin exchange in Ag-He collisions occurs much faster than in K-He collisions, suggesting that it may be advantageous to perform SEOP experiments with Ag atoms to reduce the timescale for the production of hyperpolarized $^3$He nuclei \cite{RMP,Baranga}.

\begin{figure}[t]
	\centering
	\includegraphics[width=0.43\textwidth, trim = 0 0 0 0]{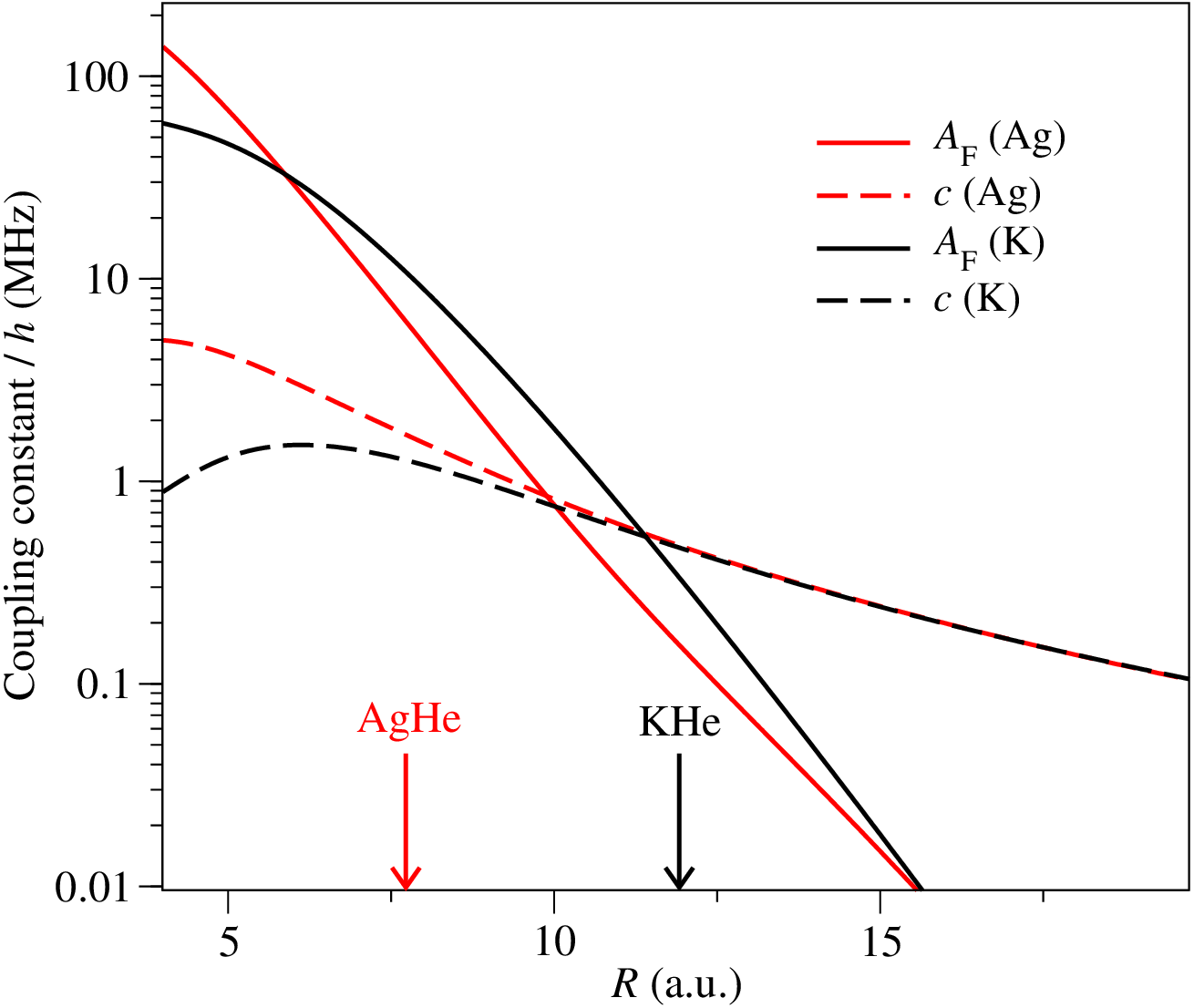}
	\renewcommand{\figurename}{Fig.}
	\caption{Spin-dependent interactions for K-He (black lines) and Ag-He (red lines) vs $R$. Full lines -- isotropic (Fermi contact) interaction constants $A_\text{F}(R)$, dashed lines -- anisotropic hyperfine constants $c(R)$. The arrows mark zero-energy turning points of the K-He and AgHe interaction potentials.}\label{fig:interactions}
\end{figure}

The Hamiltonian for the collision complex formed by a $^2S$ atom ($M$) and He may be written using atomic units \cite{RMP,pra09}
\begin{multline}\label{H}
\hat{H} = -\frac{1}{2\mu R}\frac{\partial^2}{\partial R^2}R + \frac{ \hat{\mathbf{l}}^2 }{2\mu R^2} + \hat{H}_\text{as} + \gamma(R)\hat{\bf{l}}\cdot \hat{\bf{S}} +V(R) \\ + A_\text{F}(R)\hat{\mathbf{I}}_\mathrm{He} \cdot \hat{\mathbf{S}}  + \frac{c(R)\sqrt{6}}{3}\left(\frac{4\pi}{5}\right)^{1/2} \sum_{q} Y^\star_{2q}(\hat{R})[\hat{\mathbf{I}}_\text{He}\otimes\hat{\mathbf{S}}]^{(2)}_q,
\end{multline}
where $\mu$ is the reduced mass of the complex, $\hat{\mathbf{S}}$ is the electron spin of $M$, $\hat{\mathbf{I}}_\text{He}$ is the nuclear spin of $^3$He,  $R$ is the internuclear separation, $\hat{\mathbf{l}}$ is the angular momentum for the collision, and $\hat{R}$ describes the orientation of the complex in the lab frame with the quantization axis defined by the magnetic field vector $\mathbf{B}$. In Eq. (\ref{H}), $\hat{H}_\text{as}$ is the asymptotic Hamiltonian, which describes non-interacting collision partners \cite{pra08,pra09}. 
In contrast to previous theoretical studies \cite{pra08,pra09}, our Hamiltonian explicitly includes the radial dependence of the anisotropic hyperfine interaction (the last term in Eq. \ref{H}). The spin-rotation interaction given by $\gamma(R)\hat{\bf{l}}\cdot \hat{\bf{S}}$ does not affect the spin polarization of $^3$He, and will be excluded from our analysis. The spin-dependent interactions relevant for spin-exchange $M$-He collisions include the Fermi contact interaction $A_\text{F}(R)$, and the anisotropic hyperfine interaction parametrized by the constant $c(R)$.

\begin{figure}[t]
	\centering
	\includegraphics[width=0.43\textwidth, trim = 0 0 0 0]{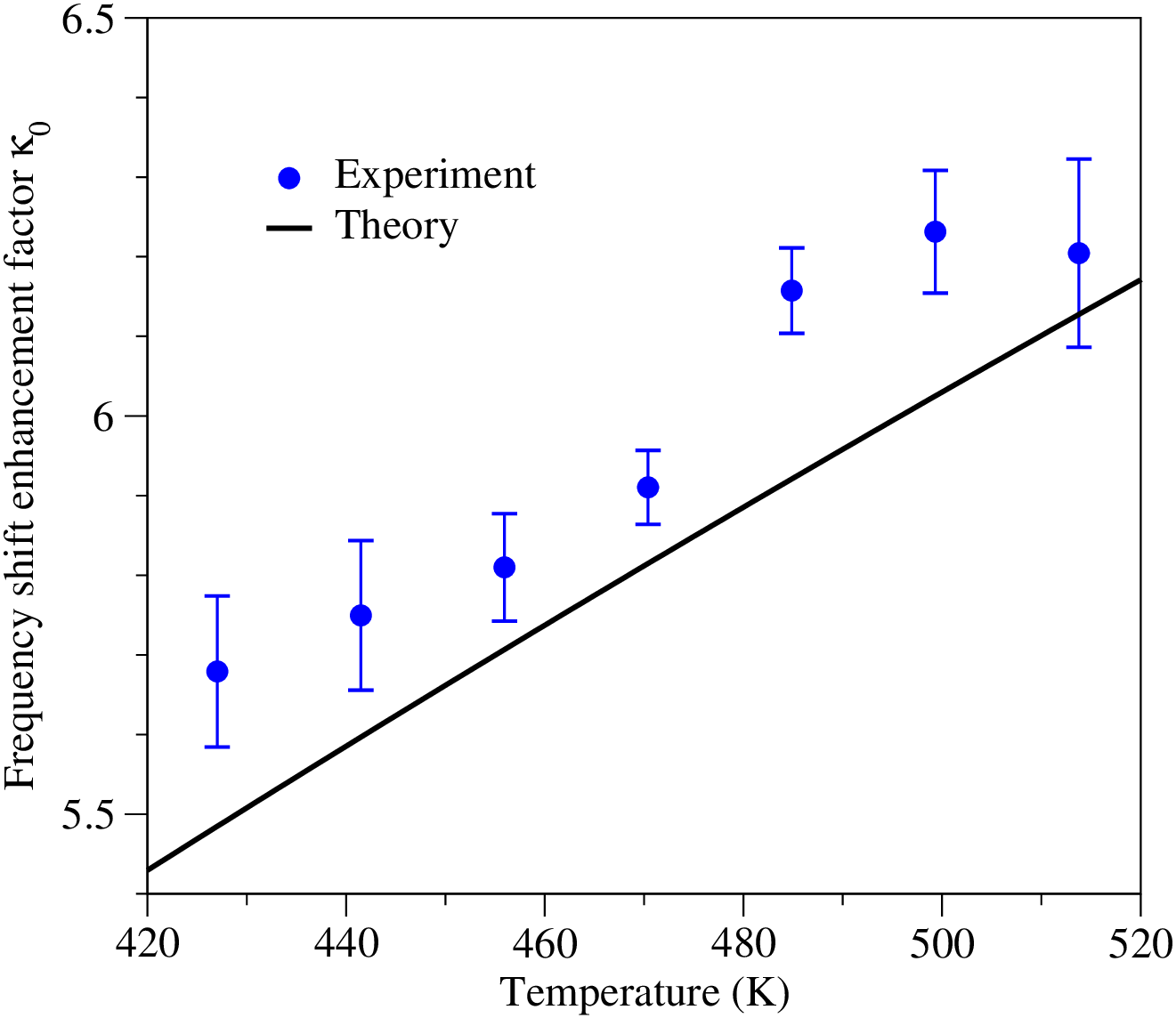}
	\renewcommand{\figurename}{Fig.}
	\caption{Frequency shift enhancement factors for K-He vs temperature: experimental data from Ref. \cite{Walker05}, present {\it ab initio} calculations (full line). }\label{fig:enhancement_factors}
\end{figure}

For the K-He interaction potential, we used the most recent, highly accurate {\it ab initio} results \cite{Partridge} fitted to analytic functions with proper long-range behavior \cite{pra09}.  We evaluated the hyperfine interaction constants $A_\text{F}(R)$ and $c(R)$ using the coupled-cluster method based on an unrestricted Hartree-Fock reference wave function \cite{UCC} as implemented in the CFOUR package \cite{CFOUR}. For the K atom, we constructed a large uncontracted basis set $(24s16p5d4f2g)$ from $(21s16p5d4f2g)$ primitives \cite{BasisK} by adding a sequence of three very tight $s$ functions with exponents forming a geometric progression. For the He atom, we employed a modified augmented correlation-consistent valence quintuple zeta (aug-cc-pV5Z) basis \cite{BasisHe} obtained by fully decontracting the $s$-functions and adding a sequence of three tight $s$ functions in  the same manner as for the K atom. To construct the Ag-He interaction potential, we fitted the {\it ab initio} data points \cite{Cargnoni} to the analytic form \cite{pra09} using the accurate long-range dispersion coefficients for AgHe \cite{Hossein}. The hyperfine constants for AgHe were evaluated using quasi-relativistic density functional theory \cite{AgHe}. We estimate the accuracy of our interaction potentials and hyperfine interactions to be $<$10\%.

In order to determine the cross sections for spin exchange in atom-He collisions, we expand the total wave function of the collision complex in basis functions $|Fm_F\rangle |I_\text{He} M_{I_\text{He}}\rangle |\ell m_\ell\rangle$, where $|\ell m_\ell\rangle$ are the partial wave states, $F$ is the total angular momentum of the atom ($F=2$ for $^{39}$K and 1 for $^{107}$Ag) and $m_F$ is the projection of $F$ on the magnetic field axis. The resulting system of close-coupled (CC) Schr{\"o}dinger equations is solved numerically on a radial grid extending from $R_\text{min}=3a_0$ to $R_\text{max}=60 a_0$ with a grid step of 0.04$a_0$. The anisotropic hyperfine interaction induces couplings between basis functions with different $\ell$ and $m_\ell$, which increases the number of CC equations to be solved for $M=0$ from 16 to 1918, where $M=m_F+M_{I_\text{He}}+m_\ell$ is the projection of the total angular momentum of the $M$-He complex on the magnetic field axis.

Our {\it ab initio} results for the isotropic and anisotropic hyperfine interaction constants for K-He and Ag-He are shown in Figure 1 as a function of the interatomic separation. Both interactions are of the order of 10 MHz for Ag-He and 1 MHz for K-He in the classically-allowed regions indicated by the vertical arrows in Fig.~1. The isotropic hyperfine interaction decreases exponentially with $R$ \cite{RMP}, and is much larger for Ag-He than for K-He as a consequence of the interaction potential for K-He being more repulsive at short $R$. The  anisotropic hyperfine interaction does not exhibit such a dramatic dependence on $R$, and is of comparable magnitude in both K-He and Ag-He complexes. This difference has important consequences for the mechanisms of spin exchange and spin relaxation in K-He and Ag-He collisions, as described below.

\begin{figure}[t]
	\centering
	\includegraphics[width=0.43\textwidth, trim = 0 0 0 0]{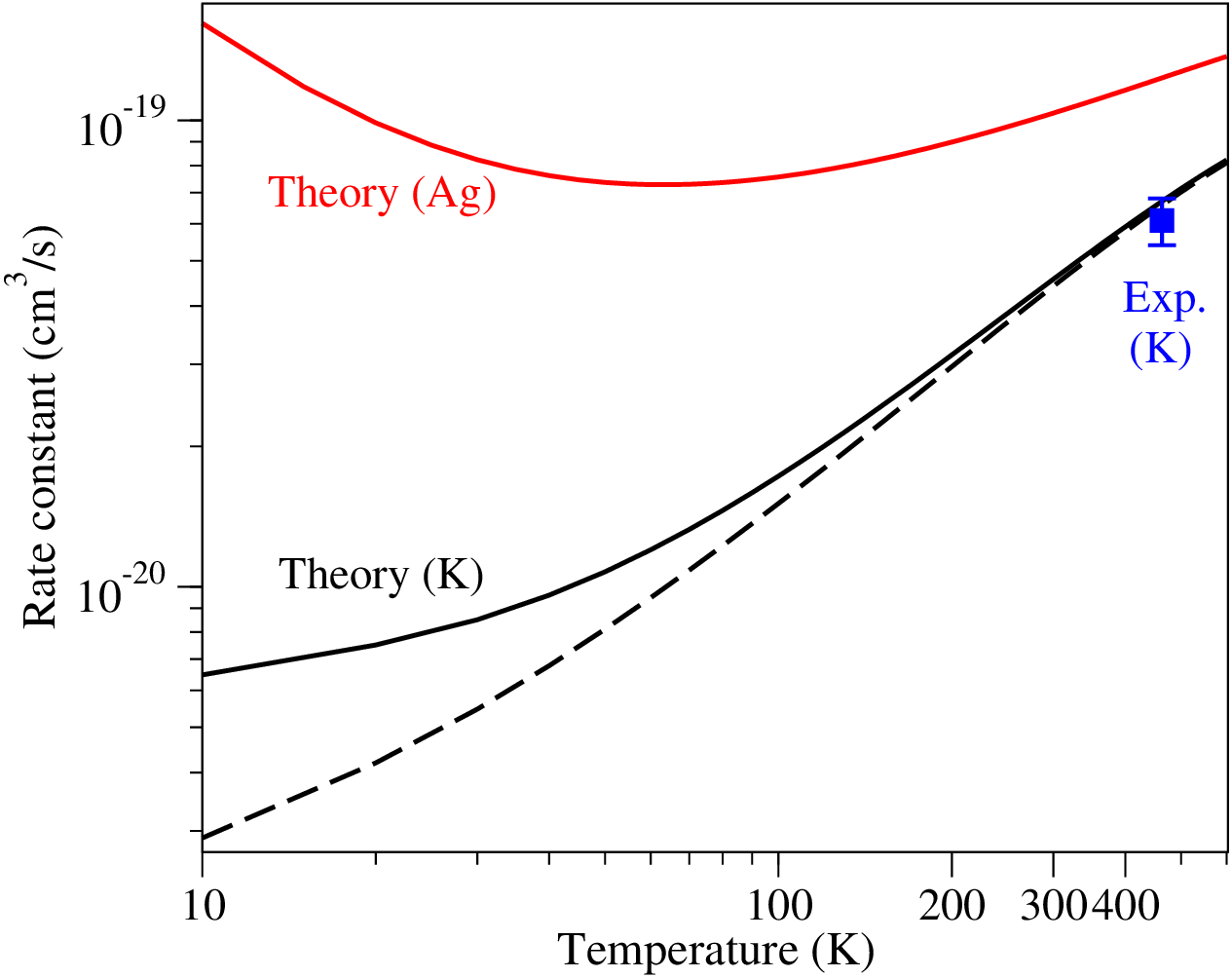}
	\renewcommand{\figurename}{Fig.}
	\caption{Rate constants for spin exchange in K-$^3$He collisions vs temperature at $B=1$ G: $k_\alpha$ (dashed line), $k=k_\alpha+k_\beta$ (full line); symbol -- experimental result \cite{Walker07,Walker10}. Also shown (red/light grey line) is the calculated $k(T)$ for Ag-He.}\label{fig:rates}
\end{figure}

To verify our {\it ab initio} interaction potentials and hyperfine interactions, we calculated the frequency shift enhancement factors 
\begin{equation}\label{Kappa}
\kappa_0(T) = \int_0^\infty \varrho_\text{He}(R) e^{-V(R)/k_BT}4\pi R^2 dR,
\end{equation}
where $\varrho_\text{He} = \frac{3I_\text{He}}{16\pi \mu_0 \mu_\text{He}}A_\text{F}(R)$  is the electron spin density of the K-He complex at the He nucleus \cite{RMP,pra09}. Since $\kappa_0(T)$ is exponentially sensitive to $V(R)$, experimental measurements of $\kappa_0(T)$  can be used as a sensitive probe to assess the quality of {\it ab initio} interaction potentials and hyperfine interactions. Figure 2 compares the calculated $\kappa_0(T)$ with the highly accurate polarimetry measurements \cite{Walker05}. Our results agree with experiment to within 3\% over the whole range of temperatures, providing an independent verification of the high accuracy of our {\it ab initio} calculations.

Figure 3 shows the total rate constant for spin exchange $k = k_\alpha+k_\beta$ in K-He collisions as a function of temperature. To facilitate comparison with experimental measurements, we distinguish between the rate constants for $^3$He spin exchange induced by the isotropic ($k_\alpha$) and anisotropic ($k_\beta$) hyperfine interactions \cite{Walker06,Walker10}. Specifically, we define $k_\alpha$ as the rate constant for the transition $|F m_F\rangle \otimes |M_{I_\text{He}}=-1/2\rangle \to |F' m_F'\rangle \otimes |M_{I_\text{He}}'=1/2\rangle$, summed over all energetically accessible final states $|Fm_F\rangle$, and we define $k_\beta$ as the rate constant for $^3$He nuclear spin depolarization $|F m_F\rangle \otimes |M_{I_\text{He}}=1/2\rangle \to |F' m_F'\rangle \otimes |M_{I_\text{He}}'=-1/2\rangle$. We note that in the absence of the anisotropic hyperfine interaction, the fully spin-polarized state of $^3$He is collisionally stable, {\it i.e.} $k_\alpha=0$.   The calculated spin exchange rate for Ag-He is notably larger than for K-He, and displays a non-monotonous  variation with temperature, increasing dramatically at $T<50$ K. This result may be qualitatively explained by the large magnitude of the Fermi contact interaction in Ag-He  as compared to K-He (see Fig. 1).

Our calculations for K-He yield  $k=6.7 \times 10^{-20}$ cm$^3$/s at 463.15~K, in quantitative agreement with the measured value of $(6.1\pm 0.7) \times 10^{-20}$ cm$^3$/s \cite{Walker10}. The anisotropic contribution to the total spin-exchange rate amounts to $1.27\times 10^{-21}$ cm$^3$/s or 1.9~\%, demonstrating that the effect of anisotropic hyperfine interaction on K-He spin exchange is weak. From Fig. 3, we observe that the rate constant $k_\alpha$ starts to deviate significantly from $k$ at temperatures below $\sim$200 K, and becomes too low by a factor of $\sim$4 at 10 K.  This observation suggests that the anisotropic hyperfine interaction has a dramatic effect on K-He collision dynamics at low temperatures, as expected based on an asymptotic analysis of Eq. (\ref{H}): in the limit of large $R$, $A_F(R)$ decreases exponentially, but $c(R)$ approaches zero as $R^{-3}$ \cite{Walter}.

\begin{figure}[t]
	\centering
	\includegraphics[width=0.43\textwidth, trim = 0 0 0 0]{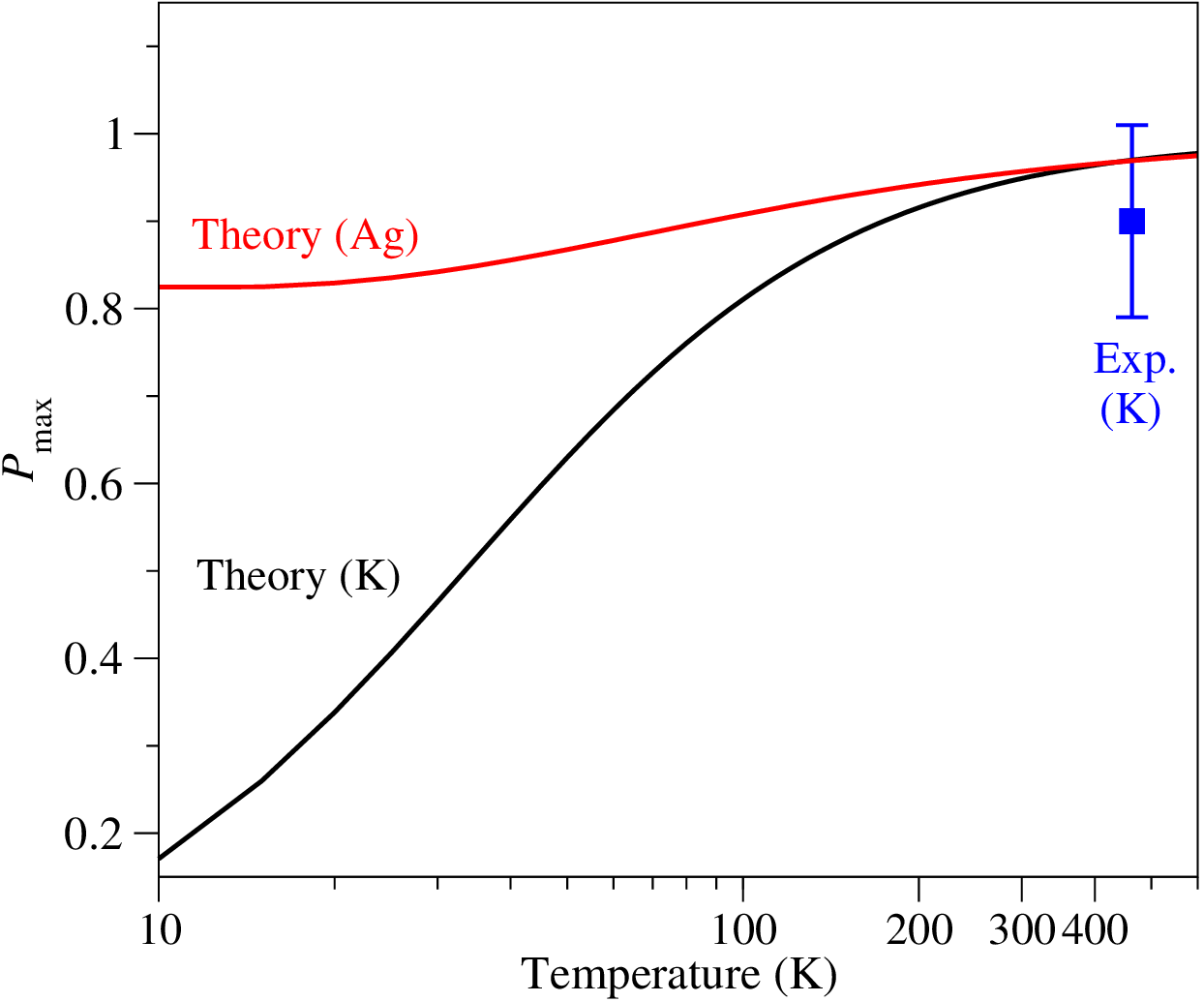}
	\renewcommand{\figurename}{Fig.}
	\caption{Maximum attainable $^3$He polarization as a function of temperature at $B=1$ G: full line -- present calculations for K-He; symbol -- experimental result for K-He \cite{Walker10}, red/light grey line -- present calculations for Ag-He. } \label{fig:rates}
\end{figure}

To elucidate the role of the anisotropic hyperfine interaction in low-temperature K-He collisions, we extended our calculations further into the cold regime and evaluated the cross sections and thermal rate constants for K electron spin depolarization induced by collisions with $^3$He atoms at $T=320$ mK and $B=1$~T.  This rate constant can be measured experimentally by observing collision-induced loss of spin-polarized K atoms from a magnetic trap \cite{pra08}. The calculated spin depolarization rate is $1.1\times 10^{-20}$ cm$^3$/s, consistent with the measured upper limit of $1.0\times 10^{-18}$ cm$^3$/s \cite{pra08}. If the anisotropic hyperfine interaction is neglected, the calculated spin depolarization rate is $0.6\times 10^{-21}$ cm$^3$/s, more than an order of magnitude smaller than the exact result.  We conclude that the anisotropic hyperfine interaction, neglected in all previous theoretical analyses \cite{pra08,pra09}, is the dominant spin relaxation mechanism of spin-polarized K  atoms trapped in the presence of cold $^3$He gas \cite{pra08}. This conclusion implies that measurements of trap loss rates of alkali-metal atoms in the presence of cold He gas may be used to probe anisotropic spin-exchange interactions in alkali-metal-He collisions.

Figure 4 shows the temperature dependence of the polarization factor $P_\text{max}=(k_\alpha - k_\beta/2)/k$, which determines the maximum attainable polarization of $^3$He that can be achieved in SEOP experiments. A falloff of $P_\text{max}$ with decreasing temperature reflects the increasing role of the anisotropic hyperfine interaction shown in Fig. 3. At $T=463.15$ K, we obtain $P_\text{max}=0.97$, in agreement with the experimental value of $0.90 \pm 0.11$ \cite{Walker10}.  The temperature dependence of $P_\text{max}$ for Ag-He  is almost identical to that for K-He at $T>400$ K, but differs significantly from the latter at lower temperatures. Since the anisotropic hyperfine interaction in Ag-He collisions is much weaker than in K-He (Fig. 1), the calculated $P_\text{max}$ for Ag-He remains high over the whole temperature range, in marked contrast to the declining behavior observed for K-He.  Figure~4 demonstrates that the efficiency of SEOP can be improved by raising the system temperature, which would further diminish the contribution due to the anisotropic hyperfine interaction and increase $P_\text{max}$. The calculated value of $P_\text{max}(T)$ increases from 0.97 to 0.98 as the temperature is varied from 463.15 to 600 K. 

In summary, we have presented a rigorous theoretical analysis of K-He and Ag-He collisions demonstrating that the maximum spin polarization of $^3$He is fundamentally limited by the anisotropic hyperfine interaction. Our calculations are in good agreement with highly accurate experimental measurements of frequency shift enhancement factors and rate constants for spin exchange in K-He collisions. Our results bear implications for research in several areas of physics with hyperpolarized $^3$He nuclei. First, they demonstrate that the maximum $^3$He spin polarization of 81\% achieved so far with SEOP experiments can be significantly improved. Second, our calculations suggest that performing SEOP experiments with atomic Ag as a collision partner will increase the SEOP rate by a factor of two at temperatures above 400 K (Fig. 2). As shown in Fig. 4, the maximum attainable He polarization via Ag-He collisions does not fall dramatically with decreasing temperature, suggesting the possibility of performing SEOP experiments in the lower-temperature regime, where the SEOP efficiency is higher due to the suppression of spin-destruction collisions driven by the spin-rotation interaction \cite{Baranga,pra08}.   Third, our analysis suggests an alternative approach to probing anisotropic spin-dependent interactions by measuring collision-induced loss of alkali-metal atoms from a magnetic trap in the presence of cryogenic He gas at milli-Kelvin temperatures. Finally, our study provides accurate reference information on the rate constants for spin-exchange and spin depolarization over a wide range of temperatures, which could be used to better constrain the anomalous spin-dependent interactions between nucleons \cite{Kimball,Petukhov} and quantify the sources of noise and decoherence in atomic magnetometers due to spin depolarization in K-He collisions \cite{Romalis}.

We thank T.G. Walker for encouraging discussions. This work was supported by the DOE Office of Basic Energy Science and NSF grants to the Harvard-MIT CUA and ITAMP at Harvard University and the Smithsonian Astrophysical Observatory.


\end{document}